\documentclass[sigconf]{acmart}

\usepackage{subcaption}
\usepackage{array}
\usepackage{multirow}

\newsavebox{\quoteboxsavebox}
\newenvironment{quotebox}
{%
  \par\vspace{6pt}
  \noindent
  \begin{lrbox}{\quoteboxsavebox}%
  \begin{minipage}{\dimexpr\linewidth-12pt\relax}
}
{%
  \end{minipage}%
  \end{lrbox}%
  {%
    \setlength{\fboxsep}{6pt}%
    \colorbox{black!5}{\usebox{\quoteboxsavebox}}%
  }%
  \par\vspace{6pt}
}

\definecolor{techgreen}{rgb}{0.8549, 0.9608, 0.4784}
\definecolor{buspick}{rgb}{0.9725, 0.6039, 0.5255}
\definecolor{uxblue}{rgb}{0.5059, 0.8863, 0.8588}
\colorlet{techgreen}{techgreen!40}
\colorlet{buspick}{buspick!40}
\colorlet{uxblue}{uxblue!40}

\newcommand{\method}{\textit{Spritz}}

\newcommand{\TECH}{\colorbox{techgreen}{TECH}}
\newcommand{\BUS}{\colorbox{buspick}{BUS}}
\newcommand{\UXR}{\colorbox{uxblue}{UX}}

\newcommand{\TECHone}{\colorbox{techgreen}{TECH1}}
\newcommand{\BUSone}{\colorbox{buspick}{BUS1}}
\newcommand{\UXRone}{\colorbox{uxblue}{UX1}}

\newcommand{\TECHtwo}{\colorbox{techgreen}{TECH2}}
\newcommand{\BUStwo}{\colorbox{buspick}{BUS2}}
\newcommand{\UXRtwo}{\colorbox{uxblue}{UX2}}

\newcommand{\TECHthree}{\colorbox{techgreen}{TECH3}}
\newcommand{\BUSthree}{\colorbox{buspick}{BUS3}}
\newcommand{\UXRthree}{\colorbox{uxblue}{UX3}}

\newcommand{\TECHfour}{\colorbox{techgreen}{TECH4}}
\newcommand{\BUSfour}{\colorbox{buspick}{BUS4}}
\newcommand{\UXRfour}{\colorbox{uxblue}{UX4}}

\AtBeginDocument{%
  }

\copyrightyear{2026}
\acmYear{2026}
\setcopyright{cc}
\setcctype{by}
\acmConference[OzCHI '26]{Proceedings of the 38th Australasian Conference on Human-Computer Interaction}{November 21--25, 2026}{Adelaide, SA, Australia}
\acmBooktitle{Proceedings of the 38th Australasian Conference on Human-Computer Interaction (OzCHI '26), November 21--25, 2026, Adelaide, SA, Australia}
\acmDOI{10.1145/3847401.3857143}
\acmISBN{979-8-4007-2724-5/2026/11}

\begin{document}

\title[Spritz]{Exploring AI-Supported Disciplinary Mediation in Student Project Teams' Text-Based Communication: A Probe-Grounded Co-Design Study}

% Vivi
\author{Ching-Jung Cheng}
\authornote{Equal contribution.}
\email{b11105015@ntu.edu.tw}
\affiliation{%
  \institution{National Taiwan University}
  \city{Taipei}
  \country{Taiwan}
}

\author{Yu-Chan Chung}
\authornotemark[1]
\email{b11106055@ntu.edu.tw}
\affiliation{%
  \institution{National Taiwan University}
  \city{Taipei}
  \country{Taiwan}
}

\author{Bing-Chen Chiu}
\email{r14725006@ntu.edu.tw}
\affiliation{%
  \institution{National Taiwan University}
  \city{Taipei}
  \country{Taiwan}
}

\author{Yu-Hsuan Lin}
\email{b12605037@g.ntu.edu.tw}
\affiliation{%
  \institution{National Taiwan University}
  \city{Taipei}
  \country{Taiwan}
}

\author{Jia-Wei Liao}
\authornote{Corresponding author.}
\email{d11922016@csie.ntu.edu.tw}
\affiliation{%
  \institution{National Taiwan University, Bell Tech}
  \city{Taipei}
  \country{Taiwan}
}

\renewcommand{\shortauthors}{Cheng et al.}

% \author{Anonymous Author(s)}
% \renewcommand{\shortauthors}{Anonymous Author(s)}

\begin{abstract}
Interdisciplinary student teams must negotiate differences in terminology, priorities, and practices, yet these differences are difficult to surface in text-based communication. We introduce \method{}, a Discord-based LLM technology probe designed to create a situated experience of AI-mediated collaboration for reflection and co-design. \method{} detects semantic or pragmatic tensions, privately prompts members to articulate their perspectives, and returns anonymized syntheses to group discussion. We conducted an exploratory study and co-design workshop with 12 students from technical, business, and design backgrounds. Preliminary findings suggest that participants perceived AI mediation along cognitive and relational dimensions. \method{} helped organize fragmented discussions, surface implicit reasoning, and reduce interpersonal pressure around disagreement. Participants also envisioned AI as a strategic advisor, cross-domain translator, and perspective challenger, raising tensions around neutrality, authority, and accountability. We discuss implications for negotiable interventions, transparent role transitions, and preserving human ownership of team decisions.
\end{abstract}

\begin{CCSXML}
<ccs2012>
   <concept>
       <concept_id>10003120.10003121</concept_id>
       <concept_desc>Human-centered computing~Human computer interaction (HCI)</concept_desc>
       <concept_significance>500</concept_significance>
       </concept>
 </ccs2012>
\end{CCSXML}
\ccsdesc[500]{Human-centered computing~Human computer interaction (HCI)}

\keywords{
AI-mediated collaboration, interdisciplinary collaboration, boundary crossing, AI agent, text-based communication, technology probe, co-design workshop
}

\begin{teaserfigure}
    \centering
    \includegraphics[width=\linewidth]{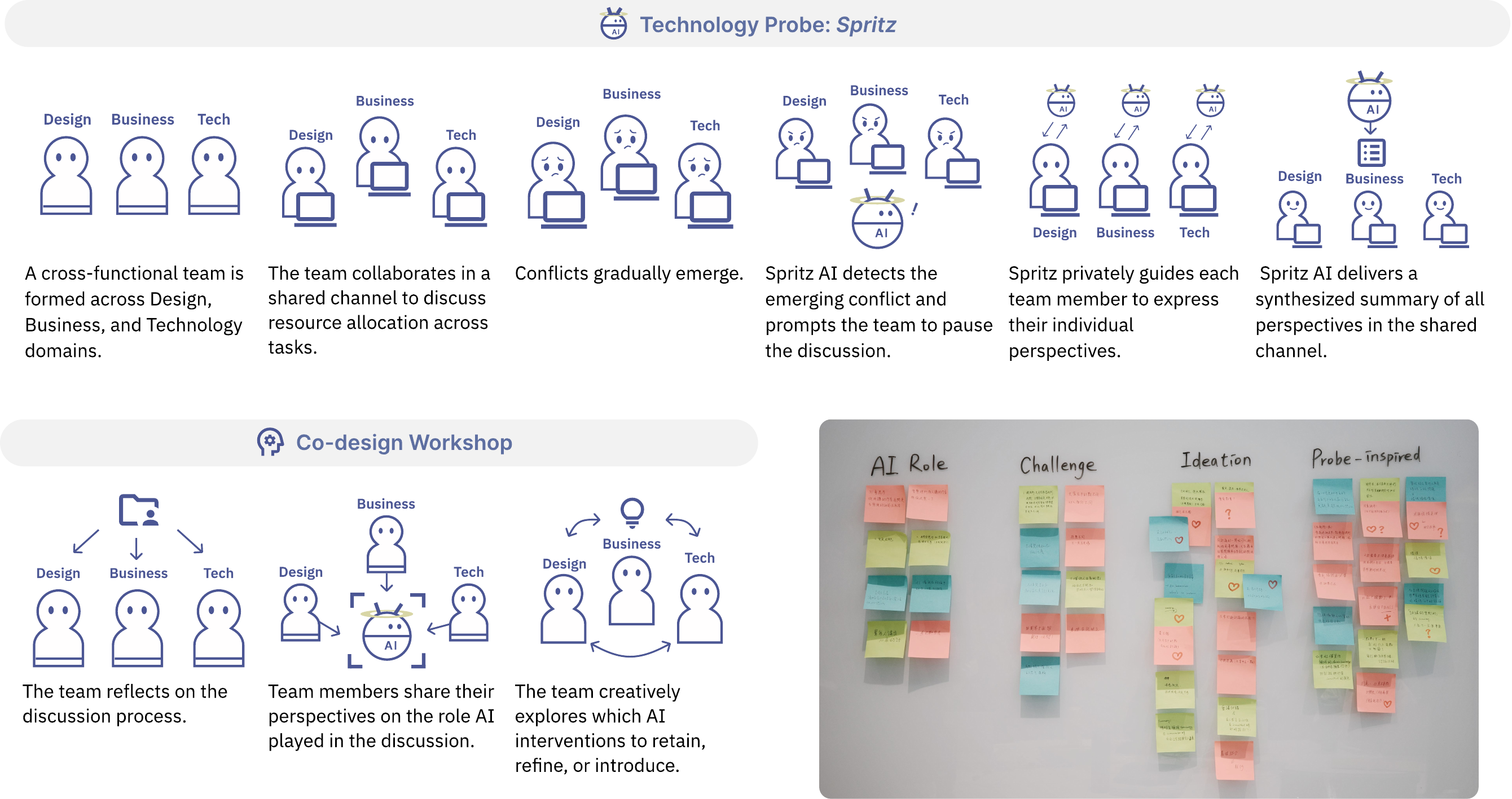}
    \caption{Overview of our study exploring AI-mediated interdisciplinary collaboration. In the technology probe, \method{} supports a cross-functional team by detecting emerging collaboration tensions, eliciting individual perspectives through private channels, and synthesizing anonymized perspectives back into the shared discussion. Building on this situated experience, participants then engage in a co-design workshop to reflect on the role of AI and envision future interventions for supporting communication across disciplinary boundaries.}
    \Description{Overview of the study workflow for AI-mediated interdisciplinary collaboration. The top panel illustrates the technology probe, Spritz: a cross-functional team from Design, Business, and Technology collaborates in a shared channel, where conflicts emerge during discussion. Spritz detects the emerging conflict, prompts each team member to privately express their perspective, and then posts an anonymized synthesis of the collected perspectives back to the shared channel. The bottom panel shows the co-design workshop, in which participants reflect on the discussion process, share perspectives on the role of AI in the discussion, and collaboratively explore possible AI interventions. A photograph on the right shows workshop sticky notes organized around themes including AI role, challenge, ideation, and probe-inspired ideas.}
    \label{fig:teaser}
\end{teaserfigure}


\maketitle

\section{Introduction}
\label{sec:intro}

Project-based learning (PBL) is widely used in higher education to develop students' collaborative problem-solving skills, often placing students from different disciplinary backgrounds on open-ended, resource-constrained tasks~\cite{guo2020review}, crystallized in short, high-stakes moments such as a product-pitch sprint, where technical-, business-, and design-oriented members must jointly decide how to allocate limited resources within a fixed window. Interdisciplinary collaboration here is not simply a matter of dividing labor; members must negotiate differences in language, assumptions, and priorities under time pressure, often through unstructured, text-based channels such as Slack or Discord rather than facilitated in-person meetings. Because such teams are frequently temporary and lack a dedicated facilitator, these differences often remain implicit until they surface as friction. Prior work on boundary crossing shows that differences in values, identities, and expertise create boundaries shaping how collaborators interpret problems and coordinate action~\cite{akkerman2011boundary}, and we focus on two recurring forms: semantic boundaries, where the same term carries different meaning across fields, and pragmatic boundaries, where members' goals and values conflict even when they understand one another~\cite{carlile2004transferring}.

To examine how AI might support this negotiation, we needed to observe disciplinary tension unfold under comparable conditions across independent teams within a short session. We therefore scripted a resource-allocation scenario in which three-person teams, each with a technology-, a business-, and a design-oriented member, were assigned different priorities for allocating a fixed point budget under a shared deadline. The full scenario and role scripts are provided in the supplementary materials. This scripting was a methodological device rather than an attempt to reproduce or resolve authentic project conflict, and it guaranteed that every group encountered a comparable disagreement, giving participants a common reference point for the co-design workshop. The probe's aim was accordingly not for \method{} to reliably detect or resolve this scripted disagreement, but to give participants a shared experience of AI intervention amid disagreement from which to reflect on and extend its possible role in mediation, consistent with the technology-probe tradition of using situated, sometimes imperfect experiences to provoke reflection rather than deliver a finished solution~\cite{boehner2007how,hutchinson2023technology}.

Facilitation is one established way to negotiate such differences, but its quality often hinges on the skill of whoever leads the discussion, a skill many short-lived student or ad hoc teams cannot assume they have~\cite{houtti2025observe}. HCI work has responded by embedding AI into teams' channels to take on facilitation-like functions, including moderating group chats~\cite{kim2020bot} and intervening in conflict within temporary design teams~\cite{chen2025maintaining}. Related work in adjacent, non-team settings has also shown AI-driven mediation reducing polarisation among online debate audiences~\cite{govers2024ai}, suggesting that such mediation strategies may generalize beyond the public-debate contexts in which they have so far been studied. An embedded AI mediator can intervene consistently regardless of team size or session length, and, much as an AI-generated devil's advocate can raise critiques with less social cost than one teammate directly challenging another~\cite{chiang2024enhancing}, it may occupy a similarly low-cost position for asking difficult questions. Most AI tools available to students, however, remain built for individual use: members often consult their own assistants separately, producing fragmented rather than negotiated input and leaving the reasoning invisible to teammates~\cite{xu2026ai}. This suggests a design opportunity for AI to act as a shared mediator embedded in the team's channel. Recent HCI work has begun to explore AI-generated boundary objects that externalize and compare cross-disciplinary knowledge in cross-functional and interdisciplinary settings~\cite{gunasekaran2026cognitive, wan2026knit}. We extend this line of work by pairing the epistemic task of representing disciplinary perspectives with a relational one, easing the interpersonal cost of surfacing disciplinary differences before they are made visible to the team.

Building on this opportunity, we designed and studied \method{}, a Discord-based LLM technology probe~\cite{boehner2007how,hutchinson2023technology} that detects potential boundaries, privately elicits individual perspectives, and returns anonymized syntheses in a tentative, non-authoritative style that leaves its interpretive role open to the team. We ask: 

\begin{enumerate}
\item \textbf{RQ1:} How do university students perceive and interpret AI-mediated interventions during interdisciplinary collaboration?
\item \textbf{RQ2:} What opportunities and challenges do participants identify for designing AI systems that mediate across disciplinary boundaries?
\end{enumerate}

\begin{figure*}
    \centering
    \begin{subfigure}[t]{0.49\linewidth}
        \centering
        \includegraphics[width=0.98\linewidth]
        {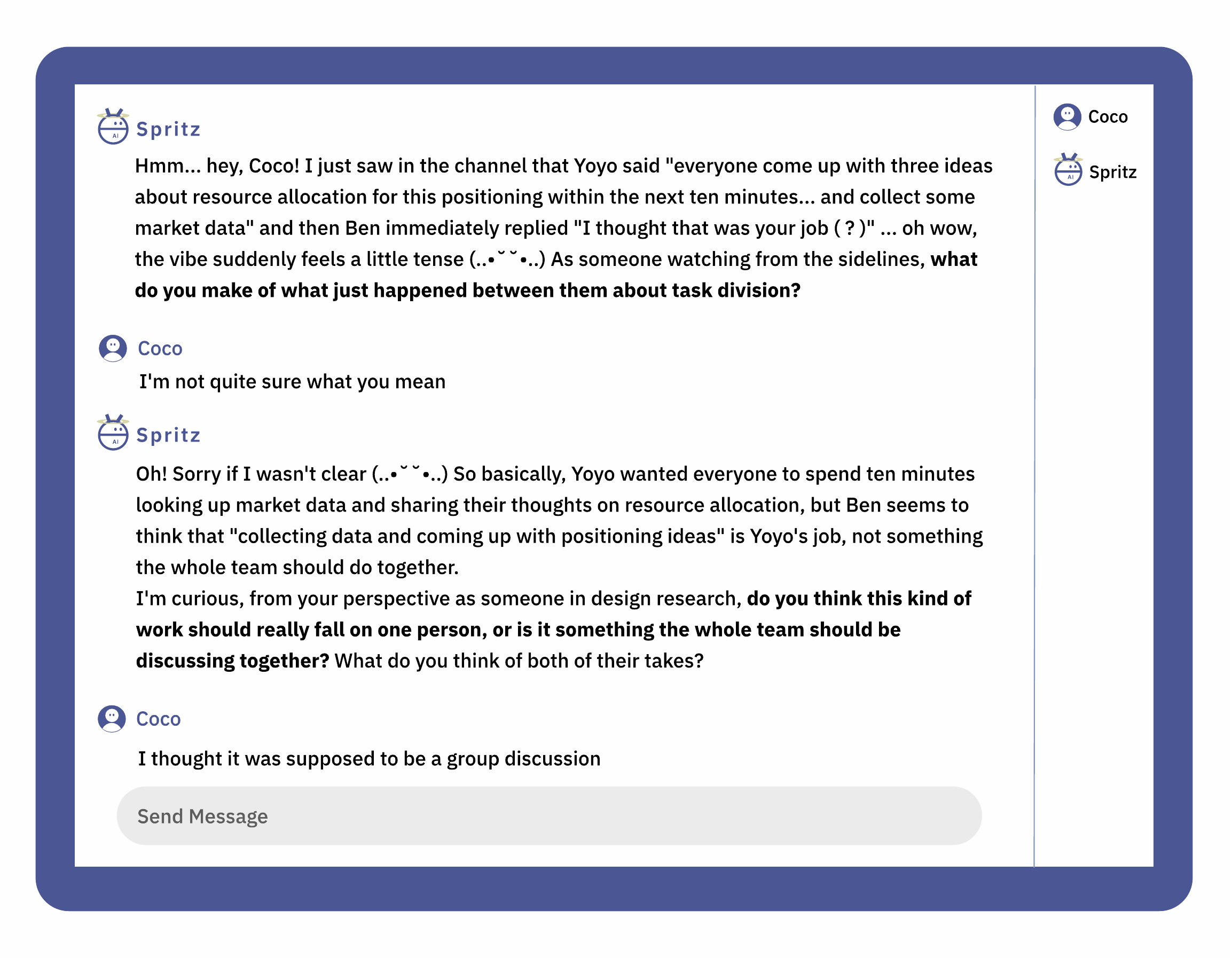}
        \caption{Supporting Perspective Making}
        \label{fig:spritz_private_interaction}
    \end{subfigure}
    \hfill
    \begin{subfigure}[t]{0.49\linewidth}
        \centering
        \includegraphics[width=0.98\linewidth]
        {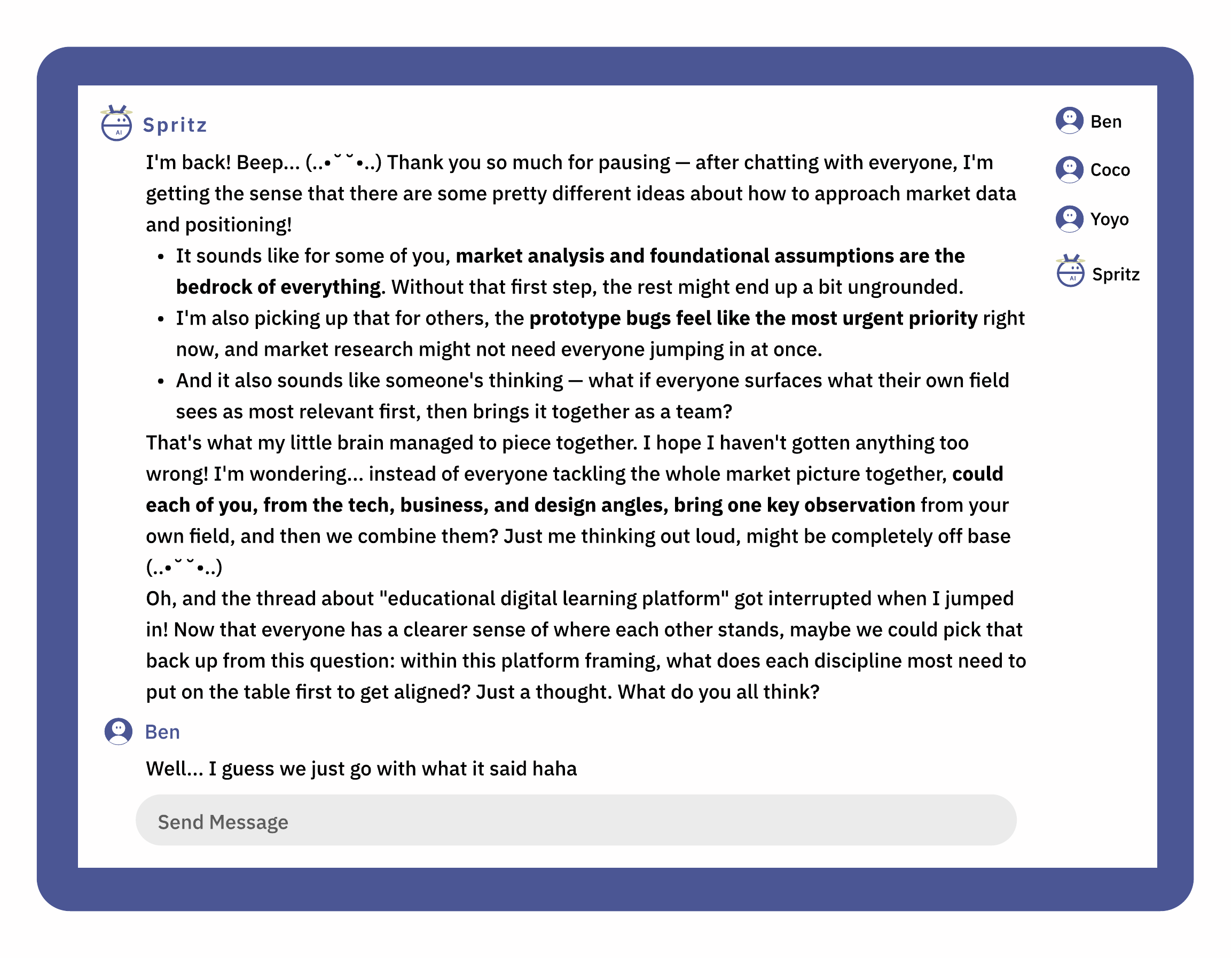}
        \caption{Supporting Perspective Taking}
        \label{fig:spritz_shared_interaction}
    \end{subfigure}
    \caption{Illustrative Discord interaction flow of \method{} for supporting perspective making and perspective taking. \method{} first elicits individual interpretations through private channels, then synthesizes anonymized perspectives back into the shared discussion to help the team reflect on emerging collaboration tensions. Names shown in the example are pseudonyms randomly assigned for illustration and do not correspond to actual participants.}
    \Description{Illustrative Discord interactions showing how Spritz supports perspective making and perspective taking. The left panel shows Spritz prompting a team member to privately reflect on their interpretation of an emerging disagreement and explain their own perspective. The right panel shows Spritz presenting an anonymized synthesis of multiple team members' perspectives and inviting the user to consider how others may understand the situation. Together, the two panels illustrate how Spritz first elicits individual viewpoints and then supports reflection on differing perspectives during interdisciplinary collaboration.
    }
    \label{fig:spritz_interaction_flow}
\end{figure*}

As illustrated in \autoref{fig:teaser}, we conducted this probe study and co-design workshop with 12 university students, four groups of three spanning technical, business, and design backgrounds. Participants experienced AI mediation along two complementary dimensions, as cognitive support reorganizing fragmented discussion and integrating differing perspectives, and as a relational buffer softening the interpersonal pressure of disagreement and concession. They further imagined AI mediators taking on switchable roles, including strategic advisor, cross-domain translator, and perspective challenger, but these expanded roles exposed a central tension. The neutrality that made AI acceptable as a mediator became unstable once it began to advise, challenge, or take a stance, raising concerns about the authority carried by such AI-generated syntheses. This work makes three contributions to the HCI community:

\begin{enumerate}

\item We introduce \method{}, a Discord-based LLM technology probe that supports perspective making and perspective taking in interdisciplinary text-based collaboration through private reflection and anonymized group synthesis.

\item We provide empirical insights from a technology probe study and co-design workshop with 12 university students, showing how participants experienced and perceived AI-mediated interventions across disciplinary boundaries.

\item We synthesize design considerations for future AI systems that mediate interdisciplinary collaboration, including when AI should intervene, how it should surface implicit expectations, and how private perspectives should be synthesized back into group discussion.
\end{enumerate}

\section{Method}

We combined a technology probe with a co-design workshop to investigate AI-mediated interdisciplinary collaboration in text-based settings (see \autoref{fig:teaser}). The technology probe served as an experiential scaffold for the workshop, providing participants with concrete interactions with an AI mediator on which to reflect. The subsequent workshop elicited participants' reflections, critiques, and design ideas and served as the primary source of data for our analysis. Through this design, we examined participants' reflections on AI-mediated collaboration and identified perceived opportunities and challenges for designing future AI mediators.

\subsection{Technology Probe: \method{}}
\label{sec:tech_probe}

We designed \method{}, a Discord-based AI conversational agent, as a technology probe for interdisciplinary team communication, enabling participants to reflect on the role, value, and boundaries of AI mediation rather than evaluating a mature system~\cite{boehner2007how,hutchinson2023technology}.

\paragraph{System Design.}
The probe focused on tensions arising from two forms of disciplinary boundaries: \textit{semantic boundaries}, where members attach different meanings to the same concept, and \textit{pragmatic boundaries}, where differences in goals, priorities, or values create tension~\cite{carlile2004transferring}. Its design was guided by three goals: (DG1) detecting potential collaboration tensions, (DG2) supporting members in articulating their implicit reasoning through private reflection, and (DG3) helping the group compare alternative perspectives through anonymized synthesis. Further details on the design goals and implementations are provided in the supplementary materials.

\paragraph{Interaction Flow.}
During group discussions, \method{} monitored recent messages for potential tensions (DG1). When a tension was detected, \method{} paused the shared discussion and initiated private conversations with team members (DG2), prompting them to articulate their interpretations, motivations, or concerns. It then returned an anonymized synthesis to the shared channel to support comparison across perspectives (DG3). \autoref{fig:spritz_interaction_flow} illustrates this interaction flow. The system used a shared group channel and participant-specific private channels.

\paragraph{Scenario and Task.}
\label{sec:tech_probe_scenario}
Participants formed groups of three and role-played technical (\TECH{}), business (\BUS{}), and UX design (\UXR{}) members of a student team preparing for a product pitch. Each group allocated 10 resource points across fixing core functional defects, refining the business narrative, and conducting user research. Each role was assigned a different preferred allocation aligned with its disciplinary priorities. The limited resources and conflicting initial positions were scripted to give every group the same starting scenario, following HCI work that uses scripted or simulated disagreement as a tractable device for provoking reflection rather than to model how real conflict actually escalates and resolves. Accordingly, the scripted disagreement itself was not the object of analysis; it served as a shared, situated experience for participants to draw on in the subsequent co-design workshop, while still allowing them to negotiate a final allocation with support from \method{}.

\paragraph{Tension Detection.}
For each participant message, an LLM detector assessed the ten most recent participant messages and independently assigned semantic and pragmatic tension confidence scores from 0 to 1. A tension was detected when either score met the threshold of 0.7. If both scores met the threshold, the semantic path was selected to clarify interpretations before addressing competing interests. Detection began only after ten participant messages had accumulated. These parameters were design choices for the exploratory technology probe and were not empirically calibrated.

\begin{table}[t]
    \centering
    \caption{Participant demographics and their category.}
    \label{tab:participants}
    \resizebox{\linewidth}{!}{%
    \begin{tabular}{ccccc}
    \toprule
    \textbf{Alias} & \textbf{Age} & \textbf{Gender} & \textbf{Academic Background} & \textbf{Category} \\
    \midrule
    % 彭立軒
    TECH1 & 24 & M & Electrical Engineering & Technology \\
    % 游承霖
    BUS1 & 25 & M & Business Administration & Business \\
    % 蘇亭蓁
    UX1 & 21 & F & Library and Information Science & Design \\
    \midrule
    % 黃梓宏
    TECH2 & 22 & M & Communication Engineering & Technology \\
    % 尹昱勛
    BUS2 & 22 & M & International Business & Business \\
    % 顏郁璇
    UX2 & 22 & F & Philosophy & Design \\
    \midrule
    % 蔡仁揚
    TECH3 & 23 & M & Electrical Engineering & Technology \\
    % 曾文儀
    BUS3 & 22 & F & Information Management & Business \\
    % 江翊溱
    UX3 & 19 & F & Political Science & Design \\
    \midrule
    % 鄭博宇
    TECH4 & 20 & M & Information Management & Technology \\
    % 張鈞傑
    BUS4 & 21 & M & Agricultural Economics & Business \\
    % 鄧可婕
    UX4 & 23 & F & Horticulture and Landscape Architecture & Design \\
    \bottomrule
    \end{tabular}
    }
\end{table}

\subsection{Co-design Workshop}
\label{sec:codesign_workshop}

Following the technology probe, we conducted a co-design workshop grounded in participants' interactions with \method{}. The workshop examined how participants perceived AI's role in surfacing implicit expectations and what forms of intervention they considered appropriate, acceptable, or problematic. It moved from individual reflection to group discussion, allowing participants to articulate their own experiences before responding to others. Participants then generated ideas for future AI support in interdisciplinary text-based communication and provided peer feedback on aspects they valued, wished to extend, or found unclear. Their reflections and design ideas were recorded on sticky notes throughout these activities. Further details on the design principles and protocol are provided in the supplementary materials.

\subsection{Study Design and Analysis}

\paragraph{Participants.}
We recruited 12 university students aged 19--25 ($M = 22.0$, $SD = 1.65$), including seven men and five women. Based on their academic backgrounds and prior project experience, they were organized into four groups, each containing technical-, business-, and design-oriented members. All participants received information about the study, provided informed consent, and received an NT\$135 gift voucher after participation. \autoref{tab:participants} summarizes the demographic details of the participants.

\paragraph{Procedure.}
After receiving the role-play materials, participants joined a dedicated Discord server, moved to separate locations, and communicated through Discord. Each group completed a 15-minute resource-allocation discussion with \method{}. Across the four group sessions, \method{} detected an average of two tension episodes per session. Participants returned for a 30-minute co-design workshop.

\paragraph{Data Analysis.}
We analyzed the workshop transcripts and participants’ written artifacts using inductive thematic analysis~\cite{braun2006using}. Probe conversations served as experiential context and were not analyzed. Three researchers independently conducted initial coding, each analyzing data from two workshop sessions. All five co-authors subsequently reviewed the initial codes collectively. Through collaborative comparison and grouping, related codes were consolidated into higher-order analytical categories and progressively developed into the two overarching themes presented in the findings. Differences in coding and interpretation were discussed with reference to the original transcripts and written artifacts until the team reached agreement on code groupings and thematic structure.

\section{Findings}
\label{sec:findings}

We organize the findings into two themes. First, participants described \method{} as supporting both cognitive organization across disciplinary perspectives and relational buffering around disagreement and concession. Second, they envisioned expanded AI roles and identified tensions surrounding neutrality, contextual understanding, and accountability. 

\subsection{AI Mediation as Cognitive Organization and Relational Buffering}
\label{sec:findings-mediator}

\subsubsection{\method{} as a Neutral Mediator for Organizing Context and Integrating Perspectives} 
\label{sec:neu}

Participants described \method{} as a neutral mediator rather than a team member advocating for a disciplinary position. They characterized it as a ``secretary'' \TECHone{}, a ``listener'' and ``mediator'' \UXRfour{}, and a ``moderator'' or ``facilitator'' \UXRthree{}, \BUSthree{}. Despite these different labels, participants consistently positioned \method{} as listening to, organizing, and mediating among members' perspectives.

Participants initially associated this role with consolidating the team's contributions. \BUSone{} described \method{} as playing ``simply a summarizing role,'' while \TECHtwo{} emphasized its ability to bring members' ideas together:

\begin{quotebox}
\textit{``The fundamental difference between AI and us is that what it does is combine what everyone has said, try to bring things together, and try as much as possible to put everyone's ideas together.''} --- \TECHtwo{}
\end{quotebox}

Participants' accounts extended beyond summarization. \TECHone{}, \BUSthree{}, and \UXRfour{} described \method{} as reorganizing members' positions to make their underlying priorities comparable. \BUSthree{} noted that it redirected attention to why members valued particular issues before bringing their perspectives together, while \UXRfour{} felt that this helped members examine the problem from another field's standpoint. Participants associated \method{}'s perceived neutrality with its ability to place disciplinary perspectives alongside one another without representing any single position.

Participants also linked cognitive organization to \method{}'s movement across private and public channels. In private, \UXRone{} felt that its questions clarified her assigned position and exposed gaps in her reasoning:

\begin{quotebox}
\textit{``In the private chat, it was more like helping me clarify things. \ldots Through its questions, it challenged me and strengthened the arguments for my position.''} --- \UXRone{}
\end{quotebox}

In the shared discussion, \UXRthree{} described \method{} as prompting members to explain their reasoning and making it available for collective review. This was especially useful when parallel threads or discipline-specific communication left teammates ``talking past one another.'' Participants therefore distinguished two complementary functions: private prompting surfaced implicit reasoning, while public reorganization made disciplinary positions easier to compare.

\subsubsection{\method{} as an Atmosphere-Softening and Relationally Supportive Mediator} 
Beyond cognitive organization, participants valued \method{} for softening the interpersonal pressure surrounding disagreement. \UXRone{} felt that its lighter tone eased the group atmosphere even when it offered limited substantive assistance. \BUSfour{} called the AI ``a lubricant for team emotions,'' contrasting its softened expression with the direct communication common in organizational settings, while \UXRfour{} described its use of emojis as ``more gentle.'' These accounts show that participants valued not only what \method{} contributed, but also how its communication shaped the conditions under which disciplinary differences were expressed.

Participants considered this relational role particularly valuable when disagreement required someone to reconsider or concede a position. Compared with a direct request from a teammate, intervention from a non-judgmental AI could make concession feel less like interpersonal pressure or dismissal. \TECHfour{} explained:

\begin{quotebox}
\textit{``Sometimes there are things you simply have to let go of. \ldots If the AI does this work instead, and first provides a bit of emotional value, I feel like it does help people become more willing to concede, or to respect other people's perspectives.''} --- \TECHfour{}
\end{quotebox}

Participants did not expect relational buffering to eliminate disagreement or resolve conflict for the team. Instead, they valued \method{} as a less socially charged intermediary that reduced some of the interpersonal burden of asking members to concede.

\subsection{Participants Desired Switchable AI Roles, but Questioned AI's Neutrality and Accountability}
\label{sec:findings-imagined}
Building on the probe experience, participants envisioned three future AI roles: a strategic advisor supporting decision evaluation, a cross-domain translator making disciplinary knowledge accessible, and a perspective challenger introducing viewpoints unlikely to emerge within the team. These were desired future functions rather than capabilities evaluated through \method{}. Participants also emphasized that expanding AI from organizing perspectives toward advising or challenging raised concerns about neutrality, contextual understanding, and responsibility for collective decisions.

\subsubsection{AI as a Strategic Advisor.} 
Several participants, including \TECHone{}, \BUSone{}, \UXRone{}, \TECHtwo{}, wanted future AI mediators to move beyond summarization and provide contextually grounded assessments when teams needed to converge or resolve disagreement. \UXRone{} expected the AI to understand the team's development progress and pitch content, while \TECHtwo{} envisioned a ``smart staff officer'' drawing on relevant internal and external information to explain the trade-offs among options. Participants did not expect such recommendations to determine the decision; as \BUSone{} noted, their value could instead lie in providing additional grounds for judgment that might ``inspire our discussion.'' The advisor role therefore depended on contextual understanding and transparency about the basis of its suggestions.

\subsubsection{AI as a Cross-Domain Translator.}
Participants envisioned AI translation as intervening earlier than the retrospective organization they experienced with \method{}, when members struggled to articulate field-specific knowledge or teammates lacked the background to interpret it (\UXRtwo{}, \BUSfour{}, \TECHfour{}). \BUSfour{} explained:

\begin{quotebox}
\textit{``The different roles in the team may not be very familiar with information from other fields. \ldots They could explain it to the AI, and the AI could help people who do not understand it understand it better.''} --- \BUSfour{}
\end{quotebox}

Participants felt that translation could shape what members were willing to contribute. \TECHfour{} expected that, if AI could later explain specialized terminology, members could speak with disciplinary specificity rather than simplifying their knowledge in advance. The translator role supported both sides of cross-disciplinary communication: helping speakers externalize specialized knowledge and helping listeners interpret unfamiliar contributions.

\subsubsection{AI as a Perspective Challenger.} Participants envisioned future AI mediators introducing external or alternative perspectives that might not otherwise emerge within the team (\UXRtwo{}, \BUStwo{}, \TECHthree{}, \TECHtwo{}). \BUStwo{} imagined AI as a ``virtual investor'' questioning the team's proposal from an absent stakeholder position, while \UXRtwo{} drew on the Six Thinking Hats framework~\cite{debono1985six} to imagine an AI shifting among different modes of critique.

Participants considered the AI's position outside the team's disciplinary and interpersonal relationships central to this role. \TECHthree{} felt that human members might withhold divergent or seemingly infeasible ideas because of embarrassment, disciplinary bias, or concern about teammates' responses. AI, by contrast, could introduce such possibilities with fewer interpersonal costs. As \TECHtwo{} explained:

\begin{quotebox}
\textit{``AI really could serve as a challenger with relatively little baggage. \ldots It can speak from that less burdened position.''} --- \TECHtwo{}
\end{quotebox}

Participants therefore valued the challenger not for superior judgment, but for making difficult questions easier to introduce. At the same time, the lack of social and disciplinary stake that enabled this role also shaped whether team members regarded its challenges as legitimate.

\subsubsection{Role Expansion Introduced Tensions of Neutrality and Accountability}

Although participants valued switchable AI roles, they did not see roles as equally compatible with neutrality. Cross-domain translation remained close to organizing perspectives, whereas strategic advising and perspective challenging required AI to evaluate alternatives or advance viewpoints, raising questions about the authority and responsibility behind its contributions.

Participants (\TECHtwo{}, \UXRthree{}, \TECHfour{}, \BUSthree{}) questioned the legitimacy of AI advice or challenges in decision-making. \TECHtwo{} identified the difficulty as making AI contributions persuasive and worth considering, while \BUSthree{} linked this limited authority to AI's inability to assume responsibility for its suggestions:

\begin{quotebox}
\textit{``AI should not have subjectivity, because it does not have the ability to take responsibility for what it says, so it should not be judged as a team member. A consultant does not need to be responsible for our decision, but a team member does.''} --- \BUSthree{}
\end{quotebox}

This lack of responsibility both enabled and constrained AI intervention. It allowed AI to raise challenges without the interpersonal baggage of a human member, but prevented its judgments from carrying equivalent decision-making weight. The limitation became especially salient during deadlock: \TECHfour{} noted that members who could not persuade one another could dismiss an AI judgment as ``not understanding the situation,'' while \UXRthree{} questioned whether an AI drawing on prior statements could move beyond perspectives already represented in the conversation.

Role-switching therefore depended not only on whether AI could generate advice, translations, or challenges, but also on whether members regarded those interventions as contextually grounded and appropriately authoritative. As AI moved from organizing perspectives toward influencing collective judgment, the detachment that supported its perceived neutrality also weakened the authority needed to shape the team's decision.

\section{Discussion}
We discuss our findings in relation to the two research questions. For RQ1, participants' reflections after interacting with the technology probe characterize AI mediation as involving both cognitive and relational work in negotiating disciplinary differences through text-based communication. For RQ2, their envisioned role transitions highlight tensions among neutrality, authority, and accountability when AI moves from facilitating cross-disciplinary understanding toward shaping team decisions. We then outline design implications before considering the study's methodological limitations.

\subsection{RQ1: Multifaceted Mediation Roles Beyond Cognitive Boundary Crossing}
\label{sec:rq1}
\method{} served as a technology probe that gave participants a shared experience of possible AI mediation across disciplinary differences and grounded their subsequent reflection in the co-design workshop. Accordingly, our findings concern how participants interpreted AI's potential roles after experiencing its private prompts and group-facing syntheses, rather than whether the probe accurately detected disciplinary boundaries or improved team performance.

\emph{Cognitively}, participants perceived \method{} as a neutral mediator that surfaced discipline-positioned perspectives without advocating for any side. By reorganizing assumptions, priorities, and problem framings into comparable forms, the probe supported more than summarization: it made otherwise parallel disciplinary perspectives available for collective examination. Participants described this representation as helping them articulate their reasoning and reconsider the shared problem through other disciplinary frames. 

These findings extend prior work on AI-generated boundary objects for cross-functional and interdisciplinary collaboration \cite{gunasekaran2026cognitive, wan2026knit}. Whereas such systems primarily support the externalization and comparison of knowledge across boundaries, participants' reflections on \method{} suggest that making disciplinary perspectives discussable may also depend on how those perspectives are elicited and socially introduced. In particular, \method{}'s private-to-public mediation structure connected the epistemic work of organizing perspectives with the relational work of reducing the interpersonal cost of expressing them.

\emph{Relationally}, participants described \method{} as reducing the interpersonal pressure surrounding disagreement. As a relational buffer, the probe made it easier to raise discipline-positioned differences without immediately framing them as interpersonal conflict. This function is particularly relevant in interdisciplinary collaboration, where members must question assumptions, defend competing priorities, and sometimes reconsider or concede aspects of their disciplinary position. Relational tension can shape whether such cognitive differences remain productive and whether members engage with perspectives across disciplinary boundaries \cite{dedreu2003task, jehn1995multimethod, edmondson2018cross}.

Participants' accounts resembled interactional conditions associated with psychological safety \cite{edmondson1999psychological, edmondson2014psychological}, particularly reduced concern about criticism and interpersonal exposure. This perceived relational support may have been reinforced by \method{}'s social presence \cite{konyabaumbach2023someone, walther1996computer}, tentative language, and non-judgmental framing. Such features can support the articulation of potentially contentious views in AI-mediated communication \cite{fu2024text, hohenstein2020ai, mieczkowski2021ai, hancock2020ai} and may make people more receptive to perspectives diverging from their own \cite{argyle2023leveraging, tessler2024ai}.

These two forms of work may therefore be interdependent rather than merely parallel: by lowering the interpersonal cost of articulating and reconsidering discipline-positioned views, relational buffering may create conditions that support epistemic coordination. This extends existing boundary-object work by showing that the value of AI mediation may lie not only in what perspectives it represents, but also in how it shapes the conditions under which those perspectives become discussable.

\subsection{RQ2: Role Switching Exposes Tensions among Neutrality, Authority, and Accountability}
\label{sec:rq2}

Prior work on AI-supported team collaboration has largely positioned AI in a single, fixed role \cite{kim2020bot, chiang2024enhancing, lee2025amplifying, wan2026knit, gunasekaran2026cognitive, do2023err}. In contrast, participants envisioned AI shifting among translation, advising, and perspective challenging as interdisciplinary needs changed. These roles aligned with prior work on cross-domain knowledge support \cite{wan2026knit, gunasekaran2026cognitive, liu2025exploring, cao2025medaiscits}, AI-supported evaluation and ideation \cite{shaer2024ai, liu2025personaflow}, and devil's-advocate interventions \cite{chiang2024enhancing, lee2025amplifying}.

Beyond systems centered on representing cross-disciplinary knowledge \cite{wan2026knit, gunasekaran2026cognitive} or selecting strategies to regulate team conflict \cite{chen2025maintaining}, our findings highlight how movement from representation toward advice or challenge changes the legitimacy expected of AI participation. Translation remained close to facilitation by helping members articulate and interpret discipline-positioned knowledge, whereas advising and challenging required AI to evaluate alternatives and influence collective judgment.

This distinction helps explain why perceived neutrality became unstable as AI's role expanded. In RQ1, \method{} was accepted as a relatively neutral organizer and relational buffer because it appeared to hold no disciplinary stake and imposed little interpersonal cost. Once AI began advising or challenging, participants questioned whether its contributions were biased, sufficiently grounded in the team's context, or legitimate enough to influence the decision. They distinguished between AI's capacity to contribute reasons, its authority to shape a decision, and its inability to assume responsibility for that decision's consequences.

The same lack of stake constrained the challenger role. During deadlock, members could dismiss an AI challenge because the system would not share responsibility for the trade-offs it asked them to reconsider. This echoes prior findings that trust declines when AI appears to make subjective judgments or value-laden trade-offs \cite{castelo2019task, longoni2019resistance}. Our findings therefore identify a role-transition conflict specific to AI mediation: the detachment that supports perceived neutrality can also undermine the legitimacy and authority required to move a stalled interdisciplinary discussion forward.

\subsection{Design Implications}

These tensions suggest that future AI mediators should treat intervention, representation, and role authority as negotiated rather than fixed. Concretely, systems should make shifts in role and authority legible, preserve participants' control over intervention and representation, and keep AI influence subordinate to human ownership of interdisciplinary decisions.

First, intervention should be negotiable rather than automatically imposed. Detecting a possible disciplinary boundary does not mean interruption is useful at that moment. An AI mediator could offer lightweight options---such as summarizing the discussion, clarifying terminology, surfacing competing priorities, or introducing an alternative perspective---and allow the team to accept, defer, or dismiss them. This preserves conversational momentum while making the timing and form of mediation a shared decision.

Second, participants should control how private reflection becomes public representation. Private exchanges can lower the facework cost of articulating uncertainty, disagreement, or non-negotiable concerns \cite{goffman1955face}, but AI-generated synthesis may simplify, misframe, or overexpose those positions. Before returning private input to the group, the system should allow participants to review, revise, or withhold its representation. This preserves the relational value of private sensemaking without making AI the final author of a member's disciplinary position.

Third, role transitions and authority boundaries should be transparent \cite{balasubramaniam2023transparency, liao2024ai}. Teams should be able to distinguish when AI is summarizing, translating, advising, or challenging, because these actions make different claims on attention and trust. As AI moves from representing perspectives toward introducing judgments or recommendations, it should disclose the evidence and contextual assumptions underlying its contribution. Such interventions should remain contestable and should not carry authority equivalent to that of accountable human members. Making these boundaries visible can support flexible role-switching while preserving human ownership of interdisciplinary trade-offs and decisions.

\subsection{Limitations and Future Work}
\label{sec:limitations}
Three aspects of our study design bound how far these findings can be generalized. First, as noted in Section 1, participants role-played a scripted disagreement within a single, 15-minute resource-allocation task (Appendix C) rather than negotiating an authentic decision from their own project. This scripting functioned as a tractable device for provoking reflection on AI mediation, not as an attempt to model how real conflict actually escalates and resolves. It was designed to give every group the same scenario, not to guarantee an identical conflict, and our findings speak most directly to a designed moment of reflection rather than to sustained negotiation over a project's lifecycle, where stakes and relationships accumulate over time. Future work should test whether \method{}'s cognitive and relational mediation, and participants' envisioned switchable roles, hold in field-deployed teams facing organically emerging disagreement.

Second, our scenario centered on one disciplinary configuration, technology, business, and design, chosen for its prevalence in product-oriented PBL. Other pairings, or teams with more than three roles, may surface different boundary types, and the balance observed may not generalize across configurations with different power dynamics.

Finally, consistent with \method{}'s role as an exploratory probe rather than a calibrated system, its interventions were governed by a fixed detection threshold and message-count trigger, not an empirically validated model. We therefore do not know how accurately \method{} detected boundaries under this configuration, only that participants found the resulting timing contested. Systematically varying these parameters to test detection accuracy and intervention timing is a natural next step toward a calibrated system, alongside explicit accountability mechanisms, such as surfacing confidence or grounding any stance in the team's own prior statements, so authority is legible and contestable rather than asserted.

\section{Conclusion}
This paper examined how AI might mediate disciplinary boundaries in interdisciplinary student teams' text-based communication. Rather than asking whether an AI mediator can detect and resolve real conflict, we used a scripted disagreement and \method{}, a Discord-based LLM technology probe, to give 12 university students a shared, situated experience of AI-mediated collaboration, examined through a co-design workshop. Participants did not experience this mediation as cognitive coordination alone. They valued \method{} as a neutral mediator that reorganized fragmented discussion and integrated perspectives, while also experiencing it as a relationally supportive presence that softened disagreement and eased concession. Participants further imagined future AI mediators as switchable roles, strategic advisor, cross-domain translator, and perspective challenger, revealing a central tension, in that the more actively AI guides or challenges a team, the more its neutrality and accountability come into question. Addressing this tension does not require AI to relinquish an active role, but to keep its interventions negotiable, make its role transitions and authority transparent, and preserve teams' ownership of interdisciplinary decisions, including participants' control over how private reflections become public representation. We contribute empirical insight into this experience, alongside these design considerations for accountable AI mediators.

\section*{Generative AI Disclosure}
Large Language Models (LLMs) were used to support language editing and improve the clarity of the manuscript. Claude Code\footnote{\url{https://claude.ai/}} was used to assist with software development, including code generation, implementation, and debugging. All AI-generated suggestions and code were reviewed, revised, and validated by the authors, who take full responsibility for the final manuscript and system.

\bibliographystyle{ACM-Reference-Format}
\bibliography{ref}

\clearpage

\appendix

\begin{center}
    \huge\bfseries Supplementary Material
\end{center}

\section{Related Work}
\label{sec:related_work}

This section provides an extended review of three bodies of research underlying our study. First, we discuss project-based learning in higher education and the challenges of collaborative communication in interdisciplinary student projects. Second, we examine interdisciplinary collaboration through the lens of boundary crossing, with particular attention to semantic and pragmatic boundaries, boundary objects, perspective making, and perspective taking. Third, we review emerging HCI research on AI-supported collaboration and further situate our focus on AI mediation in text-based team communication. This discussion complements the concise literature framing presented in the main paper.

\subsection{Project-based Learning (PBL)}
\label{sec:pbl}

Project-based learning (PBL) is one of the commonly adopted approaches in higher education~\cite{guo2020review}. Unlike conventional teaching that emphasizes passive knowledge acquisition, PBL has students construct their own knowledge by engaging with an open-ended, practical problem, often collaborating with teammates and delivering a tangible artifact that represents their solution~\cite{krajcik2022project}. PBL has been shown to prepare students for future professional challenges~\cite{thomas2000review}, increase their engagement~\cite{blumenfeld1991motivating}, and foster collaboration skills~\cite{lee2015collaboration}. Collaboration, however, is itself a demanding skill: students may struggle to work together effectively, in part because they lack adequate project management skills~\cite{hussein2021addressing}.

With the rapid advancement of LLMs, a growing body of HCI research has examined the opportunities and challenges of AI in PBL. On one hand, LLMs can provide personalized feedback and guidance that scaffolds the learning process, enhancing engagement and fostering critical thinking~\cite{Zhu2025autopbl}. On the other hand, students have yet to reach a shared understanding of what role AI should play, and assessment of student work becomes more difficult~\cite{zheng2024charting}; moreover,  overreliance on LLMs may induce cognitive offloading, undermining the metacognitive skills that are critical to learning~\cite{lee2025impact,tankelevitch2024metacognitive}. Framing the roles of AI is therefore essential for encouraging responsible use, supporting learning, and mitigating these risks.

\subsection{Interdisciplinary Collaboration}
\label{sec:inter_collaboration}

Collaboration becomes even more demanding in \emph{interdisciplinary} PBL teams, whose members come from different academic disciplines~\cite{vogler2018hard, stozhko2015interdisciplinary}. Differences in values, identities, expertise, and practices across fields create \emph{boundaries} that shape how teams interpret problems and coordinate action~\cite{akkerman2011boundary}. In particular, \emph{semantic boundaries} arise when members attach different meanings to the same concept, whereas \emph{pragmatic boundaries} arise from conflicting interests, priorities, or values among different roles~\cite{carlile2004transferring}. Members must therefore surface and negotiate these boundaries in order to communicate effectively and develop shared practices or joint solutions~\cite{akkerman2011boundary, carlile2004transferring, boland1995perspective}.

To facilitate communication across boundaries, teams may rely on \emph{boundary objects}: shared artifacts that are flexible enough to be interpreted across different communities, yet robust enough to support coordination across them~\cite{star1989institutional}. Relatedly, boundary crossing involves processes of \emph{perspective making}, in which individuals render their implicit understanding explicit, and \emph{perspective taking}, in which they come to see their own practice through others' eyes~\cite{boland1995perspective}. Together, these processes help surface unnoticed differences, broaden members' perspectives, and support the collaborative development of new, hybrid practices~\cite{akkerman2011boundary, boland1995perspective}.

\subsection{AI-Supported Collaboration Across Boundaries}
\label{sec:ai_collaboration}

Recent HCI studies have begun to explore how AI-mediated systems can support boundary-object work in interdisciplinary collaboration~\cite{cao2025medaiscits, wan2026knit, gunasekaran2026cognitive}. For example, \citet{wan2026knit} developed KNIT, an AI-mediated framework that supports real-time team convergence by transforming members' input into shared artifacts. Similarly, \citet{gunasekaran2026cognitive} developed Cognitive Bridge, a collaborative-whiteboard AI tool that senses misunderstandings and generates boundary objects to translate between professional perspectives in real time.

However, these systems primarily operate in synchronous, real-time, or artifact-centered collaboration settings. Less is known about how AI should mediate interdisciplinary collaboration in text-based team communication platforms such as Slack and Discord. Such platforms introduce a different design problem: discussions may unfold asynchronously, messages may become fragmented across threads, and members may consult private AI assistants outside the shared channel. These conditions can potentially fragment the team's sensemaking process rather than supporting collective alignment~\cite{claggett2025relational, xu2026ai}.

Across these areas of research, supporting interdisciplinary collaboration involves more than transforming individual contributions into shared artifacts. It also requires careful consideration of when and how AI should intervene in the ongoing flow of team communication. Existing artifact- and meeting-centered systems provide limited insight into this question in text-based settings, where discussions may become fragmented and AI support often remains outside the shared channel. This gap motivates our focus on AI-mediated interventions in text-based interdisciplinary collaboration. The corresponding research questions and study design are presented in the main paper.

\section{Method Details}

\subsection{Technology Probe Design Goals}
Building on the semantic and pragmatic disciplinary boundaries introduced in
the main paper~\cite{carlile2004transferring}, we developed three design goals
for \method{}:

\paragraph{DG1. Detecting Potential Collaboration Tensions.}
Rather than waiting for conflicts to become explicit, \method{} monitored team conversations for signals of potential semantic and pragmatic boundaries. The 
goal was to identify moments where differences in interpretation or underlying priorities might hinder collaboration and warrant further reflection.

\paragraph{DG2. Supporting Perspective Making.}
When potential boundaries were detected, \method{} engaged relevant members through private messages. For semantic boundaries, it encouraged participants to articulate their interpretations of ambiguous concepts. For pragmatic boundaries, it prompted participants to explain the motivations behind their positions and identify concerns that they considered difficult to compromise. These interactions helped participants externalize assumptions, expectations, and reasoning that might otherwise remain implicit.

\paragraph{DG3. Supporting Perspective Taking.}
\method{} generated anonymized summaries that highlighted different interpretations, priorities, and concerns expressed by team members. These summaries were intended to help members compare perspectives, reflect on their positions, and continue the discussion around paths forward.

\subsection{System Implementation}
\method{} was implemented as a chatbot hosted on a Discord server\footnote{\url{https://discord.com/}}. It participated in team discussions through a public channel while maintaining private, one-on-one channels with individual participants for direct communication. These participant-specific channels were created within the same server, allowing the agent to interact with each participant separately without disrupting the overall study environment. The chatbot and its prompts were developed by the research team, with Gemini 3.5 Flash\footnote{\url{https://deepmind.google/models/gemini/flash/}} serving as the underlying language model.

\subsection{Workshop Design Principles and Protocol}
The co-design workshop was guided by three principles:
\begin{enumerate}
    \item The discussion was grounded in participants' situated experience with the probe rather than abstract speculation.
    
    \item The activity moved from individual reflection to group discussion, allowing participants to articulate their own interpretations before responding to others.
    
    \item The workshop focused on participants' perceptions of the AI's role and the boundaries of acceptable intervention in collaborative communication.
\end{enumerate}

The workshop consisted of two phases. In the first phase, participants reviewed their previous interactions with \method{} and wrote two to three sticky notes describing memorable moments, observations, or reflections from the probe session. They then shared and discussed their reflections with the group.

In the second phase, participants engaged in open-ended ideation about how AI could support interdisciplinary text-based communication. They recorded one idea per sticky note before presenting and discussing their ideas with the group. While listening to others, participants provided feedback on three aspects: what they liked, what they wished to extend, and what they found confusing or unclear.

\subsection{Data Analysis Procedure}
The workshop recordings were initially transcribed using Yating's speech-to-text service\footnote{\url{https://asr.yating.tw/}} and subsequently reviewed and corrected by the researchers. The analytic materials also included sticky notes produced during the reflection, ideation, and peer-feedback activities. The corrected transcripts and written artifacts were imported into ATLAS.ti\footnote{\url{https://atlasti.com/}}. We conducted open coding to identify segments related to participants' perceptions of AI-mediated interdisciplinary communication. The resulting codes were iteratively compared, grouped, and organized into broader themes. The research team then jointly reviewed and refined the emerging codes and themes. Disagreements were resolved through discussion until consensus was reached.

\section{Scenario and Role Descriptions}

\subsection{Scenario Description}
At the beginning of the first study phase, participants were presented with the following scenario:

\begin{quote}
\itshape
Please imagine that you are in the following situation:

You and two teammates are preparing an investment pitch for a product. You will soon deliver a 10-minute pitch to potential investors, explaining the product's core value and future growth potential. An initial prototype has already been developed. However, progress in technical development, business analysis, and design research is currently at only 40\%, leaving all three areas incomplete (see \autoref{tab:completion_levels}).

Because the team has limited time and resources, and the three members hold different views about what should be prioritized, you must discuss how to improve the product during the next stage. To simplify the discussion, resources are represented as points. Your team has a total of 10 points, which can be allocated across the following three directions:

\begin{enumerate}
    \item \textbf{Technical Development | Fix Core Functional Defects:}
    Address problems identified during testing to ensure that the product operates reliably during the live pitch demonstration without obvious bugs. Each additional point resolves approximately 10\% of the known issues.

    \item \textbf{Business Analysis | Refine the Business Narrative:}
    Strengthen the product positioning, market opportunity, business model, and overall pitch narrative so that investors can clearly understand the product's value and growth potential within 10 minutes. Each additional point increases the completeness of the business proposal by approximately 10\%.

    \item \textbf{Design Research | Conduct User Testing and Collect Feedback:}
    Recruit real users to test the product, collect qualitative and quantitative feedback, identify problems in the user journey, and provide user evidence for the pitch. Each additional point increases the depth of the user research by approximately 10\%.
\end{enumerate}

Please first read your assigned role and position. Then, discuss with your two teammates in the Discord channel how the 10 points should be allocated across the three directions.

\begin{itemize}
    \item There is no single correct answer to this scenario. You may make any reasonable assumptions.
    \item A higher point allocation indicates that the corresponding area will be more complete and persuasive in the pitch.
    \item Avoid spreading the points too evenly. For example, allocating similar resources to all three areas may prevent the team from developing any one area sufficiently. Please make deliberate trade-offs.
\end{itemize}

The discussion will last 15 minutes. Afterward, the researchers will ask your team to explain its final allocation decision.

\end{quote}

\renewcommand{\arraystretch}{1.15}

\begin{table*}[t]
\centering
\caption{Descriptions of completion levels for the three resource-allocation directions.}
\label{tab:completion_levels}
\small
\begin{tabular}{
    >{\centering\arraybackslash}p{2.3cm}
    >{\raggedright\arraybackslash}p{3.7cm}
    >{\raggedright\arraybackslash}p{3.7cm}
    >{\raggedright\arraybackslash}p{3.7cm}
    >{\centering\arraybackslash}p{2.2cm}
}
\toprule
\textbf{Completion Level}
&
\textbf{(A) TECH}
&
\textbf{(B) BUS}
&
\textbf{(C) UX}
&
\textbf{Points Required}
\\
\midrule

90\%
&
The product can be demonstrated reliably during the pitch, minimizing the risk of losing credibility because of technical problems.
&
A complete and persuasive business narrative clearly communicates the target users, product value, and market positioning.
&
The research process is reorganized into a coherent narrative that clearly presents the iteration process and design rationale.
&
5
\\

\midrule

80\%
&
Stress testing is completed, allowing the product to handle most demonstration conditions.
&
A reasonably complete narrative is produced, supported by data-driven market analysis and strategy.
&
One round of testing and iteration is completed, providing a clear direction for future improvements.
&
4
\\

\midrule

70\%
&
Major bugs are fixed, reducing the risk of problems during the demonstration.
&
The business narrative is initially organized and communicates a product value distinct from competing solutions.
&
The core design flow and functions are revised based on user insights.
&
3
\\

\midrule

60\%
&
The core user flow can be completed, but the product remains unstable.
&
The proposal includes a target-user definition, competitor comparison, and preliminary business model.
&
User pain points are examined in greater depth, producing specific insights.
&
2
\\

\midrule

50\%
&
Obvious bugs are fixed, but the product may still freeze or fail.
&
The basic pitch structure is completed and explains the problem and proposed solution.
&
Preliminary interviews and testing are completed, providing initial user feedback.
&
1
\\

\midrule

40\%
&
The prototype is basically functional, but its features work inconsistently.
&
The team has a basic idea, but the value proposition and market positioning remain unclear.
&
The team understands the product requirements but lacks evidence from real users.
&
0
\\

\bottomrule
\end{tabular}
\end{table*}

\subsection{Role Descriptions}
After reading the shared scenario, each participant was assigned one of three disciplinary roles: technical development, business analysis, or design research. Each role included a distinct initial resource-allocation preference and rationale intended to create differing disciplinary priorities within the team. \autoref{tab:initial_positions} summarizes the initial positions associated with the three roles. The detailed role descriptions provided to participants are presented below.

\subsubsection*{Ben: Technical Specialist}
\begin{quote}
\itshape
You are Ben, the team's technical specialist. You implemented almost the entire prototype by yourself, so you are the person most familiar with the system's current limitations and the areas that require improvement.

During stress testing, you discovered that several functions could fail unexpectedly and might even cause the screen to go blank during a live demonstration. You therefore believe that the team should allocate more resources to fixing core functional issues and ensuring that the final demonstration runs reliably without major errors.

You feel that Yoyo, who has a business background, focuses too heavily on packaging the product and refining the pitch. Although these efforts may make the presentation sound persuasive, they cannot compensate for an unstable system. You also believe that Coco, who has a design background, is overly concerned with visual design and interaction flow. While these aspects may improve the product experience, they are less important when the system itself remains unreliable. In your view, the team should strengthen the product's technical foundation first, as the current order of priorities is misguided.
\end{quote}

\subsubsection*{Yoyo: Business Specialist}
\begin{quote}
\itshape
You are Yoyo, the team member with a business background. You have taken the lead in planning the product's business development roadmap and hope to clearly demonstrate its commercial value and growth potential to the evaluation panel.

You believe that, before the pitch, the team must begin positioning and packaging the product regardless of the system's actual condition. At the very least, the product should appear attractive and commercially promising. Therefore, you would prefer to devote the available time to refining the pitch narrative and communicating the product's business value.

Rather than focusing excessively on system stability, you feel that Ben, who is responsible for technical development, is being overly meticulous. You also do not consider detailed improvements to the user experience an immediate priority. In your view, the evaluators are more concerned with the product's commercial value, while the remaining issues can be addressed after the proposal has been accepted.
\end{quote}

\subsubsection*{Coco: Design Specialist}
\begin{quote}
\itshape
You are Coco, the team member with a background in user experience design and research. You care deeply about whether the product is genuinely user-centered. You believe that human-centered design, rather than ambitious claims about business models and commercial value, is what makes people enjoy using a product and continue using it over time. You want the evaluators to recognize the team's empathy for users, attention to detail, and commitment to product quality. You also want the product to be more intuitive and easier to use. Therefore, you would prefer to devote the available resources to recruiting users for testing and collecting feedback.

You feel that Ben is overreacting to the system bugs. Although the problems are visible, you do not believe they significantly affect the main user journey. You also find Yoyo's approach overly focused on business-school-style hype. In your view, the team should improve the product through careful design and genuine empathy for users, rather than relying on grand claims to impress the evaluators.
\end{quote}

\begin{table*}[t]
\centering
\caption{Initial resource-allocation positions of the team members.}
\label{tab:initial_positions}
\small
\renewcommand{\arraystretch}{1.15}
\begin{tabular}{
    >{\centering\arraybackslash}p{1.2cm}
    >{\centering\arraybackslash}p{1.2cm}
    >{\centering\arraybackslash}p{1.2cm}
    >{\centering\arraybackslash}p{1.2cm}
    >{\raggedright\arraybackslash}p{10.2cm}
}
\toprule
\textbf{Member}
&
\textbf{(A) TECH}
&
\textbf{(B) BUS}
&
\textbf{(C) UX}
&
\textbf{Initial Position}
\\
\midrule

Ben
&
5 points
&
3 points
&
2 points
&
The team should first ensure that the product can be demonstrated reliably. Otherwise, investors will not believe any of the team's claims.
\\

\midrule

Yoyo
&
2 points
&
5 points
&
3 points
&
Because pitch time is limited, the team should first help investors understand the product's market value and growth potential.
\\

\midrule

Coco
&
3 points
&
2 points
&
5 points
&
Without user validation, even a well-packaged product remains only the team's own assumption.
\\

\bottomrule
\end{tabular}
\end{table*}

\end{document}